\documentstyle[12pt]{article}

\begin{document}

\begin{flushright}{OITS 755}\\
July 2004
\end{flushright}

\vspace*{0.6cm}

\begin{center}
{\Large {\bf Dihadron Correlation in Jets Produced \\ in
Heavy-Ion Collisions}}
\vskip .75cm
  {\bf   Rudolph C. Hwa$^1$ and  C.\ B.\ Yang$^{1,2}$}
\vskip.5cm

  {$^1$Institute of Theoretical Science and Department of
Physics\\ University of Oregon, Eugene, OR 97403-5203, USA\\
\bigskip
$^2$Institute of Particle Physics, Hua-Zhong Normal University,
Wuhan 430079, P.\ R.\ China}
\end{center}
\vskip.5cm
\begin{abstract} 
The difference between the structures of jets 
produced in heavy-ion and hadronic collisions can best be exhibited 
in the correlations between particles within those jets. We study the 
dihadron correlations in jets in the framework of parton 
recombination. Two types of triggers, $\pi^+$ and proton, are 
considered. It is shown that the recombination of thermal and shower 
partons makes the most important contribution to the spectra of the 
associated particles at intermediate $p_T$. In $pp$ collisions the 
only significant contribution arises from shower-shower 
recombination, which is negligible in heavy-ion collisions. Moments 
of the associated-particle distributions are calculated to provide 
simple summary of the jet structures for easy comparison with 
experiments.
\end{abstract}
\section{Introduction}
Recent experiments at RHIC have revealed extensive information on the effects
of the dense medium on hadron production at large transverse momentum
($p_T$) \cite{ka,dd,jw}.  The suppression of  back-to-back correlation relative
to the same-side correlation is a strong indication of substantial energy loss
suffered by hard partons propagating through the medium  \cite{ca}.   It
implies then that the jets detected in heavy-ion collisions are produced
mainly near the surface of the medium so that the hard partons are less
attenuated by the jet quenching effects \cite{xw}.   If that is 
indeed the case,
then a simple fragmentation model would predict that the structure of jets
produced in heavy-ion collisions should be basically the same as that 
of jets in
$pp$ collisions.  That similarity has been shown to be absent in the data of
more recent experiments \cite{fq}.  The aim of this paper is to investigate
the jet structure by examining the two-particle correlation within a jet
produced by nuclear collisions at high energy.  The difference from $pp$ jets
is naturally caused by the presence of thermal partons in the jet
environment, which we shall take into account.

The framework in which we shall study hadron production at large $p_T$ is
parton recombination, which has been shown to explain some features of the
data where fragmentation fails \cite{hy}-\cite{hy2}.  The anomaly associated
with species dependence of the Cronin effect in d+Au collisions has also
been resolved in the recombination picture \cite{hy3,hy4}.  The main
component that is new in this series of work is the inclusion of shower partons
generated by hard partons.  Whereas the recombination of two shower
partons forms a hadron that can be identified with the fragmentation product
in the conventional approach, the recombination of shower partons with
thermal partons yield hadrons in the intermediate $p_T$ region that are
totally new.  Here we go a step further.  To study dihadrons in a jet we must
consider four partons that recombine to form two hadrons; some of those
partons will be thermal in order to have enhanced yield.  Dihadrons formed
without thermal partons correspond to those found in jets in vacuum, and
are suppressed compared to those hadron pairs that involve the participation
of thermal partons.

We shall consider two types of triggers, pion and proton, and calculate the
distributions of the associated particles.  Since we include the contributions
from different species of hard partons, each having various flavors of shower
partons, and since the four recombining partons have different momentum
fractions that have to be permuted in the recombination formula, the
combinatorial complication can result in a hundred terms or more.  For that
reason we limit our trigger to only $\pi^+$ and $p$, which are sufficient to
reveal the properties of the jet structures.  Detailed comparison of 
our results
to current data is, however, difficult, since the experimental trigger at this
stage consists of all charged hadrons, as are the associated particles
\cite{ca,fq,jr}.  Separating triggers to mesons and baryons has not resulted in
any $p_T$  distributions for the associated particles \cite{as}.  Nevertheless,
some coarse comparisons with the data can be made, and our results will be
shown to be reasonable.

\section{Single- and Two-particle Distributions}

The hadronization process that we consider in order to calculate the dihadron
correlation in a jet is parton recombination.  The formalism for 
single-particle
inclusive distribution at high $p_T$ is given in \cite{hy2}.  An essential
ingredient in the hadronization process is the shower parton distributions
(SPD), which give the probabilities of finding shower partons of various
flavors and momentum fractions in jets initiated by different hard
partons \cite{hy5}.  Convoluting the SPD's, denoted by $S^j_i$, with the
hard-scattered parton distributions $f_i(k)$ in heavy-ion collisions gives the
shower component
\begin{eqnarray}
{\cal S}(q_1) = \xi \sum_i \int dk k
f_i(k) S^j_i (q_1/k) \ ,
\label{1}
\end{eqnarray}
where the sum is over all hard partons, and $\xi$ is the average suppression
factor due to energy loss in a dense medium, found to be 0.07 for central
Au+Au collisions at $\sqrt{s}=200$ GeV \cite{hy2}.  For two shower partons
in the same jet we have
\begin{eqnarray}
{\cal SS} (q_1, q_2)=\xi\sum_i
\int dk k f_i(k)
\left\{S^j_i\left({q_1\over k}\right),S^{j'}_i\left({q_2\over
k}\right)
\right\},
\label{2}
\end{eqnarray}
where the curly brackets denote the symmetrization of the leading parton
momentum fraction
\begin{eqnarray}
\left\{S^j_i (z_1),\ S^{j'}_i (z_2)\right\} =
{1\over  2} \left[S^j_i (z_1) S^{j'}_i \left({z_2\over 1-z_1}\right) +
S^j_i
\left({z_1\over 1-z_2}\right)S^{j'}_i (z_2)\right] .
\label{3}
\end{eqnarray}
This symmetrization is necessary, since either $j$ or $j^{'}$ shower parton
may be the leading parton in the jet, and Eq.\ (\ref{3}) is a way to 
ensure that
momentum conservation $z_1 + z_2 \leq 1$ is not violated.  The SPD's are
determined by solving the recombination formula for the fragmentation
functions (FF)
\begin{eqnarray}
  xD^M_i(x) = \int {dx_1  \over  x_1} {dx_2  \over x_2}
\left\{S^j_i (x_1),\ S^{j'}_i (x_2 )
\right\} R^{jj^{'}}_M (x_1, x_2, x) \ .
\label{4}
\end{eqnarray}
where $ R^{jj^{'}}_M$ is the recombination function (RF) for the
hadronization process $j + j^{'} \rightarrow M$.

For the hadronization of a hard parton in vacuum it is unnecessary to
consider the shower partons, since they recombine to form $M$, as in Eq.\
(\ref{4}), to recover the FF, from which the SPD's are obtained.  However, in
the environment of thermal partons as in heavy-ion collisions the shower
partons can recombine with the thermal partons, resulting in hadrons that
are dominant in the intermediate $p_T$ region because they benefit from the
high density of the thermal partons as well as the higher momenta of the
semi-hard shower partons.  The invariant distribution of the thermal partons
is parameterized by
\begin{eqnarray}
{\cal T}(q_1) = q_1{dN_q^{\rm th}\over dq_1} =
Cq_1\exp (-q_1/T),
\label{5}
\end{eqnarray}
where $T$ is the inverse slope enhanced by flow.  Since we have no model to
describe the hydrodynamical evolution of the bulk medium at low transverse
momentum, the parameters $C$ and $T$ are determined by fitting the
low-$p_T$ data.  They are found to be \cite{hy2}
\begin{eqnarray}
C = 23.2\ {\rm GeV}^{-1}, \qquad\qquad T = 0.317
\ {\rm GeV} ,
\label{6}
\end{eqnarray}
for central Au+Au collisions at midrapidity and $\sqrt{s} = 200$ GeV.

With both the thermal ${\cal T}(q_1)$ and shower distributions ${\cal
S}(q_2)$ known, it is possible to calculate the hadron distribution in the
1D recombination formalism for the formation of a meson \cite{dh,hw}
\begin{eqnarray}
p^0{dN_M  \over  dp} = \int {dq_1 \over  q_1}{dq_2
\over q_2}F_{q\bar{q}^{\prime}}  (q_1, q_2) R_M(q_1, q_2, p)\ ,
\label{7}
\end{eqnarray}
where $F_{q\bar{q}^{\prime}}  (q_1, q_2)$ is the joint distribution of a
quark $q$ at $q_1$ and an antiquark $\bar{q}^{\prime}$ at $q_2$.  For pion
production at high $p_T$, we put $\vec{p}$  in the transverse plane,
abbreviate
$p_T$ as $p$, ignore pion mass and write \cite{hy2}
\begin{eqnarray}
{dN_{\pi}  \over  pdp} =  {1 \over p^3} \int^p_0
dq_1F_{q\bar{q}^{\prime}} (q_1, p-q_1)  ,
\label{8}
\end{eqnarray}
where the RF \cite{hw,hy6}
\begin{eqnarray} R_{\pi}(q_1, q_2, p) ={q_1q_2  \over p^2}
\delta\left({q_1
\over p}+ {q_2  \over p} - 1\right),
\label{9}
\end{eqnarray}
has been used.  For heavy-ion collisions $F_{q\bar{q}^{\prime}}$ can be
written in the form
\begin{eqnarray}
F_{q\bar{q}^{\prime}} = {\cal TT} + {\cal TS} +
{\cal SS}\ ,
\label{10}
\end{eqnarray}
where the possibility of two shower partons from
two different jets is ignored.  It is the ${\cal TS}$ term in Eq.\ 
(\ref{10}) that
dominates at intermediate $p_T$, while the ${\cal TT}$ and ${\cal SS}$ terms
are dominant in the low and very high $p_T$ regions, respectively.

Two-pion distribution can be obtained by a straightforward extension of the
single-particle distribution given in Eq.\ (\ref{7}), and one gets
\begin{eqnarray}
{dN_{\pi_1 \pi_2}  \over  p_1 p_2 dp_1dp_2} =  {1 \over p^2_1 p^2_2}
\int\left(\prod^4_{i=1}{dq_i  \over  q_i}\right) F_4
(q_1, q_2,q_3, q_4) R_{\pi_1}(q_1, q_2, p_1) R_{\pi_2}(q_3, q_4, p_2) ,
\label{11}
\end{eqnarray}
where sums over different combinations of partons contributing to the two
RF's are not exhibited explicitly.  $F_4(q_1, q_2,q_3, q_4)$ is the joint
distribution of two quarks and two antiquarks.  If we are to study the two-pion
distribution in a jet, then neither of the two pions should be the
hadronization of thermal pions only.  That is, each pion should contain at
least one shower parton in order to qualify as a part of the jet.  Using the
terminology ``thermal hadrons'' to refer to the hadronization of thermal
partons only, then in the experimental analysis of jet structure such
thermal hadrons are regarded as background and are subtracted from the
set of particles associated with a trigger.  In our calculation of $dN_{\pi_1
\pi_2}/dp_1 dp_2$ associated with a jet, we simply leave out ${\cal TT}$
contribution to any pion.  Thus there are only two types of terms for $F_4$,
which we represent schematically as
\begin{eqnarray}
F_4 =  ({\cal TS}) ({\cal TS})+ ({\cal TS})({\cal SS})\ ,
\label{12}
\end{eqnarray}
where a term of the type $({\cal SS}) ({\cal SS}) $ is omitted because without
${\cal T}$ it is negligible compared to $({\cal TS})({\cal SS})$ for hadron
$p_T < 6$ GeV/c.  In Eq.\ (\ref{12}) the parentheses enclose the 
partons that are to recombine.  With that notation $({\cal TS}) 
({\cal TS})$ is very different from $({\cal TT}) ({\cal SS})$, which 
contributes to a thermal hadron that we
exclude from our consideration.

In addition to two-pion correlation we shall also study the correlation
between a pion and a proton in a jet.  The single-baryon distribution has the
general form
\begin{eqnarray}
p^0{dN_B  \over  dp} = \int\left(\prod^3_{i=1}{dq_i  \over
q_i}\right) F_3 (q_1, q_2,q_3)\ R_B(q_1, q_2, q_3, p)
\label{13}
\end{eqnarray}
where the RF for proton is \cite{hy}
\begin{eqnarray}
R_p(q_1, q_2, q_3, p) = { g_{\alpha \beta} \over  6} (y, y_2)^{\alpha + 1}
y_3^{\beta +1} \delta\left(\sum^3_{i=1}y_i -1 \right), \qquad y_i = {q_i \over
p}
\label{14}
\end{eqnarray}
\begin{eqnarray}
g_{\alpha \beta} =\left[ B (\alpha +1, \alpha + \beta + 2) B (\alpha +1,
\beta + 1)\right]^{-1}
\label{15}
\end{eqnarray}
with $\alpha = 1.75$ and $\beta = 1.05$ \cite{hy7}.  Thus the $\pi p$ joint
distribution in a high-$p_T$ jet is
\begin{eqnarray}
{dN_{\pi p}  \over  p_1 p_2 dp_1 dp_2} = {1  \over p^2_1
p^2_2} \int\left(\prod^5_{i=1}{dq_i
\over  q_i}\right) F_5 (q_1, q_2,q_3,q_4,q_5)\  R_{\pi}(q_1, q_2,
p_1) R_p(q_3, q_4, q_5, p_2) .
\label{16}
\end{eqnarray}
The 5-parton distribution $F_5$ has the schematic form
\begin{eqnarray}
F_5 =  ({\cal TS}) ({\cal TTS})+ ({\cal TS})({\cal TSS}) ,
\label{17}
\end{eqnarray}
where we have omitted terms of the type $ ({\cal TS}) ({\cal SSS})$, $ ({\cal
SS}) ({\cal TSS})$ and $ ({\cal SS}) ({\cal SSS})$ because they are 
all negligible
compared to the ones in Eq.\ (\ref{17}).

In the next section we examine in detail the multi-parton distributions $F_4$
and $F_5$ and how they contribute to the associated particle distributions,
when either the pion or the proton is used as a trigger.

\section{Distributions of Associated Particles}

To select the appropriate two-particle distributions to calculate, 
let us examine
the type of quantities that have been measured experimentally.  There exist
data from RHIC experiments that give some properties of charged particles
associated with triggers detected in certain $p_T$ ranges \cite{fq,jr,as}  We
shall therefore calculate distributions with similar kinematic ranges.  Because
of detection limitations there is at present no particle 
identification in either
the trigger or the associated particles.  We shall, nevertheless, do the
calculations for specific particle species in anticipation for the 
corresponding
data that will become available in the future.  In particular, we 
shall consider
$\pi^+$ and $p$ as trigger particles, and $\pi^{\pm}$ and $p$ as associated
particles.  Considering all charged particles would involve overwhelming
complications and uncertainties without gaining clarity.

The two-particle distributions given in Eqs.\ (\ref{11}) and (\ref{16}) do not
explicitly specify what the two particles at $p_1$ and $p_2$ are within the
same jet.  The jet momentum $k$, which is the momentum of the initiating
hard parton, is imbedded in the expression for the shower partons, Eq.\
(\ref{2}), and is integrated over all values in a collision.  When the
two-parton distribution, such as that given in Eq.\ (\ref{10}), is generalized
to $F_4$ in Eq.\ (\ref{12}), it is important to make sure that the hard partons
in each of ${\cal S}$ in (\ref{12}) are one and the same, and similarly in
Eq.\ (\ref{17}).  To make this point explicit, it is helpful to write the
two-particle distribution within one jet as
$dN^{(i)}_{h_1h_2}(k)/dp_1dp_2$, so that all shower partons are associated
with the hard parton $i$ at momentum $k$.  Then the particle $(h_2)$
distribution associated to a trigger particle $(h_1)$ at $p_1$ is
\begin{eqnarray}
\left.{ dN_{h_2} \over  dp_2}\right|_{h_1(p_1)} = { \sum_i \int dk k  f_i
(k){dN^{(i)}_{h_1h_2} \over dp_1 dp_2 }(k)  \over \sum_i \int dk k  f_i
(k){dN^{(i)}_{h_1} \over dp_1 }(k)  } \ .
\label{18}
\end{eqnarray}
Further integration of Eq.\ (\ref{18}) over $p_1$ in the specified range of the
trigger momentum yields the distribution of the associated particles that is
measured.  Note that the energy loss suppression factor $\xi$ is cancelled in
Eq.\ (\ref{18}).  In the following our expressions for ${\cal S}$  will not
contain $\xi \sum_i \int dk k f_i(k) $ that appears in Eqs.\ (\ref{1})
and (\ref{2}), since it is now shown explicitly in Eq.\ (\ref{18}).

Let us consider first $\pi^+$ trigger, and $\pi^+, \pi^-$, and $p$ associated
with it.  The initiating hard parton $i$ can be $u$, $d$, $s$,
$\bar{u}$, $\bar{d}$, $\bar{s}$ and $g$.  For every given $i$, the shower
parton $j$ can be $u$, $d$, $\bar{u}$, $\bar{d}$.  There are therefore 28
possible SPD's.  For $\pi^+\pi^+$ in a jet, $F_4$ for four partons has three
terms
\begin{eqnarray}
F^{\pi^+\pi^+}_4 =  ({\cal TS}) ({\cal TS})+ ({\cal TS})({\cal SS}) + ({\cal
SS})({\cal TS})\ ,
\label{19}
\end{eqnarray}
where the first pair of parentheses in each term correspond to the 
trigger, the second
the associated particle.  We omit the term $({\cal SS})({\cal SS})$ 
because it is
negligible.  Note that it is the only term contributing to two 
particles in a jet
produced in $pp$ collisions, since there are no thermal partons of any
significance in the environment.  Herein lies already the difference between
jets produced in heavy-ion and hadronic collisions, without even any
quantitative details to be investigated.  For $\pi^+\pi^+$, $j$ should only be
$u$ or $\bar{d}$ within each of the six sets of parentheses; the thermal
partons should just be given the flavor of the complement to make a $\pi^+$,
i.e., $u\bar{d}$.  Thus $({\cal TS})({\cal TS})$ is a symbolic short-hand for
\begin{eqnarray}
({\cal TS})({\cal TS}) = \left[({\cal T}_u{\cal S}_{\bar{d}}) +({\cal S}_u{\cal
T}_{\bar{d}}) \right] \cdot \left[({\cal T}_u{\cal S}_{\bar{d}}) 
+({\cal S}_u{\cal
T}_{\bar{d}}) \right],
\label{20}
\end{eqnarray}
where the subscript denotes the flavor.  For the other terms in Eq.\ (\ref{19})
we have
\begin{eqnarray}
({\cal TS})({\cal SS}) = \left[({\cal T}_u{\cal S}_{\bar{d}}) +({\cal S}_u{\cal
T}_{\bar{d}}) \right] ({\cal S}_u{\cal S}_{\bar{d}}) ,
\label{21}
\end{eqnarray}
\begin{eqnarray}
({\cal SS})({\cal TS}) = ({\cal S}_u{\cal S}_{\bar{d}})\left[({\cal T}_u{\cal
S}_{\bar{d}}) +({\cal S}_u{\cal T}_{\bar{d}}) \right]\ .
\label{22}
\end{eqnarray}
For $\pi^+\pi^-$ production we have in a similar way
\begin{eqnarray}
F^{\pi^+\pi^-}_4 = \left[({\cal T}_u {\cal S}_{\bar{d}}) +({\cal S}_u {\cal
T}_{\bar{d}}) \right] \cdot \left[({\cal T}_{\bar{u}}{\cal S}_d) +({\cal
S}_{\bar{u}} {\cal T}_d)+ ({\cal S}_{\bar{u}}{\cal S}_d)\right] + ({\cal
S}_u{\cal S}_{\bar{d}})\left[ ({\cal T}_u{\cal S}_{\bar{d}}) +({\cal 
S}_{\bar u}{\cal
T}_{d})\right]\ .
\label{23}
\end{eqnarray}
It should be noted that in the above expression $u$ ($\bar d$) have
momenta $q_1$ ($q_2$), and $\bar u$ ($d$) have momenta $q_3$
($q_4)$.

In the case of $\pi^+ p$ correlation there is an extra ${\cal T}$ or 
${\cal S}$,
as shown in Eq. (\ref{17}).  Hence, the 5-parton distribution has the form
\begin{eqnarray}
F^{\pi^+ p}_5 =  \left[({\cal T}_u{\cal S}_{\bar{d}}) +({\cal S}_u{\cal
T}_{\bar{d}}) \right] \cdot \left[({\cal T}_u {\cal T}_u {\cal
S}_d) + 2  ({\cal T}_u {\cal S}_u {\cal T}_d) + 2 ({\cal
T}_u{\cal S}_u{\cal S}_d) +({\cal S}_u {\cal S}_u{\cal
T}_d)\right]\ .
\label{24}
\end{eqnarray}
The momenta of the quarks in the proton are in the order 
$u(q_3)u(q_4)d(q_5)$. The two factors of 2 in the above equation 
arise from the fact that ${\cal T}_u(q_3) {\cal S}_u (q_4){\cal
T}_d(q_5)$ makes the same contribution as
${\cal S}_u (q_3){\cal T}_u(q_4) {\cal T}_d(q_5)$.  That is, the two $u$
quarks can each receive contributions from the thermal and shower sources.
The same holds for ${\cal T}_u {\cal S}_u {\cal S}_d$ and ${\cal S}_u {\cal
T}_u {\cal S}_d$ also.

For proton trigger we shall consider only $\pi^+$ and $\pi^-$ as the
associated particles.  For $\pi^+$ associated with $p$, $F^{p \pi^+}_5$ is
trivially related to $F^{ \pi^+p}_5$ simply by interchanging the positions of
the parentheses in Eq.\ (\ref{24}), i.e., interchanging trigger and associated
particle, and $p_1 \leftrightarrow p_2$.  For $\pi^-$ associated with $p$
we have
\begin{eqnarray}
F^{p \pi^-}_5 = \left[({\cal T}_u {\cal T}_u {\cal
S}_d) + 2  ({\cal T}_u {\cal S}_u {\cal T}_d) + 2 ({\cal
T}_u {\cal S}_u{\cal S}_d) +({\cal S}_u {\cal S}_u{\cal
T}_d)\right] \cdot \left[({\cal T}_{\bar{u}}{\cal S}_{d}) +({\cal
S}_{\bar u}{\cal T}_{d}) \right]\ .
\label{25}
\end{eqnarray}
The thermal partons are flavor independent, so Eq.\ (\ref{5}) can be used for
$u$, $d$ and their antiquarks.  The situation with the shower partons are far
more complicated.  Their distributions depend on the species of the initial
hard parton $i$ and the shower parton $j$.  However, by not considering the
production of kaons and hyperons, there are only three basic SPD's:  $K$,
$L$, and $G$.  We can write $S^j_i$ in the matrix form
\begin{eqnarray}
S^j_i =
\left(\begin{array}{cc}
K&L\\
L&K\\
L&L\\
G&G\\
\end{array}\right),\qquad i = u, d, s, g, \quad j = u, d.
\label{26a}
\end{eqnarray}
For example, $S^u_u = K$, $S^u_d = L$, $S^u_s = L$, $S^u_g = G$.
Antiquarks are like quarks of different flavors, i.e, $S^{\bar{u}}_{\bar{u}} =
K$,
$S^{\bar{u}}_u= L$,
$S^d_{\bar{s}}= L$, etc.  Note that hard $s$ quark is included, though not
shower $s$ quark.  The parametrizations of the SPD's are given in Ref.\
\cite{hy5}, in which one can also find a thorough discussion of why $j = g$ is
excluded due to gluon conversion.  In Eqs.\ (\ref{20})-(\ref{25}) the label $i$
for hard partons is suppressed, but is shown explicitly in Eq.\ 
(\ref{18}).  The
subscripts of ${\cal S}$ in those equations correspond to the label $j$ above.
Thus for every fixed $i$ all the ${\cal S}_j$ distributions in Eqs.\
(\ref{20})-(\ref{25}) can be rewritten as $K$, $L$ and $G$.  When $i$ is
changed, the translation also changes.  Thus, for example, when $i = u$,
$\bar{u}$, $d$, $\bar{d}$, $s$, $\bar{s}$, and $g$, the $({\cal TS})$ for
$\pi^+$ in Eq.\ (\ref{20}) becomes
\begin{eqnarray}
({\cal TS})_{i=u} = ({\cal T}L) + (K{\cal T})\ ,
\label{26}
\end{eqnarray}
\begin{eqnarray}
({\cal TS})_{i=\bar{u},d,s,\bar{s}} = ({\cal T}L) + (L{\cal T})\ ,
\label{27}
\end{eqnarray}
\begin{eqnarray}
({\cal TS})_{i=\bar{d}} = ({\cal T}K) + (L{\cal T})\ ,
\label{28}
\end{eqnarray}
\begin{eqnarray}
({\cal TS})_{i=g} = ({\cal T}G) + (G{\cal T})\ .
\label{29}
\end{eqnarray}
The index of ${\cal T}$ has been omitted because of flavor independence.  In
the case of proton, if $i = u$, then only one shower $u$ quark can be valence,
the other must be in the sea; consequently, we have for the $\cal TSS$ part in
Eq.\ (\ref{24}), for example,
\begin{eqnarray}
({\cal TSS})_{i=u} = 2({\cal T}KL) + (KL{\cal T})\ ,
\label{30}
\end{eqnarray}
\begin{eqnarray}
({\cal TSS})_{i=d} = 2({\cal T}LK) + (LL{\cal T})\ ,
\label{31}
\end{eqnarray}
while, for $i = \bar{u}, \bar{d}, s$ and $\bar{s}$, $K$'s
above are replaced by $L$'s, and for $i = g$, both $K$ and $L$ are 
replaced by $G$.

The application of Eqs.\ (\ref{26})-(\ref{29}) to  (\ref{20})-(\ref{22}) must
take into account the consideration that there can only be one valence quark
in a jet, and it can be in either the trigger or the associated 
particle.  Thus, for
example, $({\cal TS})$$({\cal TS})$ in Eq.\ (\ref{20}) for $\pi ^+\pi^+$
should have the explicit form, for $i = u$,
\begin{eqnarray}
\left[ ({\cal TS})({\cal TS})\right]_{i=u} &=& {1  \over  2}\left\{\left[({\cal
T}L)+(K{\cal T}) \right] \cdot \left[({\cal T}L)+(L{\cal T}) \right]\right.
\nonumber
\\ &&+ \left.\left[({\cal T}L)+(L{\cal T}) \right] \cdot 
\left[({\cal T}L)+(K{\cal
T})
\right]  \right\}\ ,
\label{32}
\end{eqnarray}
whereas, for $\pi^+p$, the term $ ({\cal TS})({\cal TSS})$ in Eq.\ (\ref{24})
becomes
\begin{eqnarray}
\left[ ({\cal TS})({\cal TSS})\right]_{i=u} &=& {1  \over 
2}\left\{\left[({\cal
T}L)+(K{\cal T}) \right] \cdot \left[2({\cal
T}LL)+(LL{\cal T}) \right]\right. \nonumber  \\
&&+ \left.\left[({\cal T}L)+(L{\cal T}) \right] \cdot  \left[2({\cal 
T}KL)+(KL{\cal
T})
\right] \right\}\ .
\label{33}
\end{eqnarray}

So far the complications above are due to the differences in SPD's 
for different
$i$ and $j$.  Further complications arise when the constraint due to
momentum conservation is to be applied.  Since the sum of the momenta of
shower partons in a jet cannot exceed the hard parton momentum $k$, the
momentum fractions in the arguments of the SPD's cannot be independent.
In  Eqs.\ (\ref{12}) and (\ref{17}) there are terms where ${\cal S}$ appears
twice or thrice.  Let the momenta of the two-quark case be denoted by $q_a$
and $q_b$, and let us consider the term $(K{\cal T})+(L{\cal T})$ in Eq.\
(\ref{32}) for illustration.  The constraint applies only to $K(z_a)$ and
$L(z_b)$, where $z_{a,b} = q_{a,b}/k$, while the momenta of ${\cal T}$ are
independent.  If $K$ has the leading momentum $q_a$, then the
maximum momentum of $L$ is $k - q_a$, and vice-versa.  Thus we use
the symmetrized combination given in Eq.\ (\ref{3})
\begin{eqnarray}
\left\{ K(z_a), L(z_b)\right\} \equiv  {1  \over  2} \left[ K(z_a)L\left({z_b
\over  1-z_a}
\right) + K\left({z_a
\over  1-z_b}
\right)L(z_b)\right]\ .
\label{34}
\end{eqnarray}
In the case when there are three SPD's, as in Eq.\ (\ref{33}), we symmetrize as
follows
\begin{eqnarray}
\left\{ K(z_a), L(z_b), L(z_c)\right\} \equiv  {1  \over  3} \left[ 
K(z_a)\left\{L\left({z_b \over  1-z_a}\right), L\left({z_c \over
1-z_a}\right)\right\}\right.\nonumber\\+ L(z_b)\left\{K\left({z_a
\over  1-z_b}\right), L\left({z_c \over  1-z_b}\right)\right\} \nonumber\\
\left. + \left\{K\left({z_a  \over  1-z_c}
\right), L\left({z_b \over  1-z_c}\right)\right\} L(z_c) \right]\ .
\label{35}
\end{eqnarray}
The procedure for our calculation is now completely specified.

\section{Results}

To provide an understanding of the order of magnitude of the various terms,
let us first give the result of $\pi^+\pi^+$ correlation in central Au+Au
collisions at $\sqrt{s} = 200$ GeV.  Since the STAR data \cite{ca} 
are for the trigger momentum in the range $4 < p_T < 6$ GeV/c, we 
calculate
\begin{eqnarray}
{ dN^{(\pi^+)}_{\pi^+} \over  dp_2} = \left. \int^6_4 dp_1
{dN_{\pi^+} \over  dp_2}\right|_{\pi^+(p_1)} ,
\label{36}
\end{eqnarray}
where the RHS is defined by Eq.\ (\ref{18}). The result is shown in Fig.\ 1.
The dashed line indicates the contribution from the first term in Eq.\
(\ref{19}) that has the structure  $ ({\cal TS})({\cal TS})$, while 
the dash-dot
line represents the next two terms in Eq.\ (\ref{19}) that are of the form
$ ({\cal TS})({\cal SS}) + ({\cal SS})({\cal TS})$.  The solid line 
is their sum.
Clearly, the overall distribution is dominated by the component that involves
two thermal partons due to the high density of those soft partons.  Each time
a thermal parton is replaced by a shower parton in their recombination, the
yield is lower.  For that reason we have not bothered to calculate the
contribution from $ ({\cal SS})({\cal SS})$, which corresponds to the double
application of the fragmentation function.  Presumably such contributions
can become important at very large $p_T$, where the effects of thermal parton
are insignificant.   To generate four shower partons in a jet resulting in two
pions each with $p_T > 4$ GeV/c would require a much harder collision than
is necessary if some thermal partons can participate.  The issue here is not
which channels a hard parton hadronizes into, given a value of $k$ (as one
considers in fragmentation), but rather, given two pions at $p_1$ and $p_2$,
what the most favorable  value of $k$ is  in the environment of dense thermal
partons (as one considers in recombination).

We next consider the dependence on the density of thermal partons.  Since we
have already investigated d+Au collisions in connection with the Cronin
effect \cite{hy3}, where the soft parton distributions (called thermal also)
have been determined for various centralities, we calculate $\pi ^+ \pi ^+$
correlation for three cases:  central Au+Au (0-5\%), central d+Au (0-20\%) and
peripheral d+Au (60-90\%).  The result is shown in Fig.\ 2, where the 
three cases are
represented, respectively, by solid, dashed and dash-dot lines.  Evidently, the
density of thermal partons has a crucial effect on the yield of the associated
particles.  Since peripheral (60-90\%) d+Au collisions are almost 
equivalent to $pp$ collisions,
we can see directly from Fig.\ 2 that the structures of jets in 
central Au+Au and $pp$ collisions are drastically different.  
 If we 
plot the ratio of the spectra for central Au+Au to peripheral d+Au in 
linear scale, we get the solid curve in Fig.\ 3. The ratio of central 
d+Au to peripheral d+Au is shown by the dashed line. The former ratio 
exceeds 5 around $p_T=2$ GeV/c and remains large throughout the 
intermediate $p_T$ range. We now have strong evidence that the 
structure of jets in nuclear collisions is very different from that 
in hadronic  collisions.

We now consider $\pi^+$, $\pi ^-$ and $p$ production associated with a
$\pi ^+$ trigger in the 4 to 6 GeV/c range.  Fig.\ 4 shows the three
contributions by the thin solid $(\pi^+)$, dashed $(\pi^-)$ and dash-dot
lines $(p)$, with the sum indicated by the thick solid line.  The data points
are from STAR, which includes all charged hadrons in both the trigger and the
associated particles \cite{fq}.  Since what is calculated is not exactly what
is measured, one should not expect perfect agreement.  However, $\pi^+$ is a
dominant component of the trigger, and the data average over different
trigger particles, whereas the different associated particles are summed.  Thus
what is calculated should not differ greatly from what is measured.  Indeed,
the agreement is very good both in normalization and in shape.  It is therefore
reasonable to infer that our approach has captured the essence of the physics
of hadronization.

It is interesting to note that the yield of $\pi^-$ is higher than 
that of $\pi^+$
when the trigger is $\pi ^+$, as is evident in Fig.\ 4.  The reason 
is that when
  $i$ is summed over all hard parton species, the cases when $i = d$ and
$\bar{u}$ can give rise to valence shower partons that enhance the $\pi ^-$
production through the $K$ distribution, but not the $\pi^+$ production.
When $i = u$ and $\bar{d}$, the trigger uses up the valence shower parton to
form $\pi ^+$ so the associated particle, whether $\pi ^+$ or $\pi^-$, has to
be formed by the sea shower parton through the $L$ distribution, resulting in
no big difference between $\pi ^+$ or $\pi^-$.  Thus adding up the
contributions from all hard partons results in more $\pi^-$ than $\pi ^+$ in
a $\pi ^+$ triggered jet.  This is a prediction that can be checked by
experiments with good particle identification.

We also note that the proton yield in Fig.\ 4 is greater than the $\pi ^+$
yield in the 2-3 GeV/c range because a $u$ or $d$ shower parton can
recombine with two thermal partons, thereby increasing the $p/\pi$ ratio for
the same reason that the ratio exceeds 1 without trigger \cite{hy}-\cite{hy2}.
However, the proton yield is less than the $\pi^-$ yield in the $\pi^+$
triggered jet, since $\pi^+p$ does not have the advantage of the
$\bar{u}$-initiated jet that enhances the $\pi ^+\pi^-$ production.

Finally, we come to the proton trigger and show in Fig.\ 5 the result of our
calculation for the associated particles being   $\pi ^+$ and $\pi^-$, in
dashed and dash-dot lines, respectively.  The solid line is their sum.  The
normalization and shape of the total distribution for the associated particles
are roughly the same as those in Fig.\ 4 for $\pi^+$ trigger.  The
$\pi ^+$ and  $\pi^-$ components have no noticeable difference (bearing in
mind that the $\pi^-$ curve is lowered by a factor of 2 to avoid overlap).
That is reasonable, since $i = \bar{d}$ and $\bar{u}$ favor $\pi ^+$ and
$\pi^-$  equally, while $i = u$ and $d$ are both used in the trigger, leaving
$\pi ^+$ and  $\pi^-$ again on comparable footing.  These features can also
be checked directly by experiments when particle identification is improved.

In the foregoing we have presented the distributions of the associated
particles, which offer more details than the overall yields.  The 
latter provide
a short and useful summary of the jet structure that is easier to measure.  We
therefore calculate the three lowest moments of the distributions that have
already been obtained
\begin{eqnarray}
M^{(h_1)}_n = \int ^{4.5}_{0.5}  dp_2\ p^n_2\ {dN^{(h_1)}  \over  dp_2} ,
\label{37}
\end{eqnarray}
where $dN^{(h_1)}/dp_2$ is the $p_T$  distribution of all the particles
associated with trigger $h_1$ that we have calculated:  $\pi ^+$, $\pi^-$ and
$p$ for $h_1 = \pi ^+$, and  $\pi ^+$ and  $\pi^-$ for $h_1 = p$.  The lower
limit of the integral in Eq.\ (\ref{37}) is set at 0.5 GeV/c because 
our calculated
result is not reliable for $p_T < 0.5$ GeV/c; the upper limit is set at 4.5
GeV/c, since we do not want it to exceed the average of the trigger momentum
that is between 4 and 6 GeV/c.  Thus, by definition, $M^{(h_1)}_0$ is a
measure of the average number of particles associated with trigger $h_1$,
$M^{(h_1)}_1$ being the total scalar $p_T$ of those particles,
and $M^{(h_1)}_2$ the total $p_T^2$ of them.  The last quantity is
insensitive to the low $p_2$ behavior of $dN^{(h_1)}/dp_2$, and is a good
measure for comparison between theory and experiment.  Our results on
$M^{(h_1)}_n$ for central Au+Au collisions are shown in Table I.  Note
that
\begin{table}[h]
\begin{center}
\caption{Values of the moments $M^{(h_1)}_n$ for central Au+Au
at $\sqrt{s} = 200$ GeV}
\begin{tabular}{|l||c||c||c|}\hline
  \quad  $n$&0&1&2\\ \hline\hline
$h_1=\pi ^+$&1.394&1.707&2.703\\ \hline
$h_1= p$&0.882&0.999&1.450\\ \hline
\end{tabular}
\end{center}
\end{table}
the values of the three moments change by mildly decreasing factors (0.63,
0.59, 0.54) when the trigger is changed from $\pi ^+$ to $p$.  This is a
feature that can more easily be checked by experiments than the
distributions
$dN^{(h_1)}/dp_2$ themselves.

To compare the above results with what one can expect from $pp$ collisions,
we can return to what we have already calculated, i.e., $\pi ^+ \pi ^+$
correlation in peripheral d+Au collisions, since that is very close to $pp$
collisions.  Indeed, for an appreciation of the centrality dependence we
calculate the moments of the three distributions shown in Fig. 2.  The results
are given in Table II.  The ratios
\begin{table}[h]
\begin{center}
\caption{Values of the moments $M^{(\pi ^+)}_n$ for central $\pi ^+$
associated particle only in three colliding systems}
\begin{tabular}{|l||c||c||c|}\hline
\qquad\qquad  $n$&0&1&2\\ \hline\hline
Au+Au (central)&0.428&0.498&0.754\\ \hline
d+Au (central)&0.200&0.223&0.331\\ \hline
d+Au (peripheral)&0.089&0.101&0.153\\ \hline
\end{tabular}
\end{center}
\end{table}
$R = $\ [Au+Au (central)]/[d+Au (peripheral)] are 4.8, 4.9 and 4.9 for $n = 0,
1, 2$, respectively.  They are very similar and roughly 5.  We expect 
that if all
particles associated with the trigger are included, the ratio will remain about
the same.

There are some data on total charged multiplicity and total scalar $p_T$, but
they include the trigger \cite{fq}.  The ratio for [Au+Au (central)]/$pp$ on
multiplicity is $\sim 1.3$, and on scalar $p_T$ is $\sim 1.5$.  By subtracting
out the trigger contribution, it is possible to see a rough agreement with our
result; however, we leave the quantification of the comparison to the
experiments.

\section{Conclusion}

We have studied the structure of jets produced in heavy-in collisions by
calculating dihadron correlation in the framework of parton recombination.
Since the jets are produced in the environment of dense partonic medium,
they are different from the ones produced in $pp$ collisions and $e^+e^-$
annihilation. The interaction between the hard and soft partons is very
important.  It is taken into account in our study by allowing shower partons to
recombine with the thermal partons.  Since our formalism has been applied
successfully to single-particle spectra in previous studies, we have no
freedom to adjust any part of our treatment of the two-particle distributions,
nor is there any free parameter to vary.  All the results shown in the previous
section are predictions.

In our approach to hadron production at high and intermediate $p_T$ the
effect of energy loss by the hard partons traversing the dense medium is
represented by a multiplicative factor $\xi$.  That factor is cancelled in our
definition of the associated particle distribution, which is the ratio of the
two-particle distribution to the trigger-particle distribution.  Thus the
dihadron correlation in a jet that we calculate is independent of the degree of
jet quenching.  It can be compared to the corresponding data on the
near-side jet, since such jets are produced by hard collisions near the surface
facing the detector and suffer minimal energy loss.  Unfortunately, inadequate
particle identification renders unfeasible direct comparison between the
currently available data with our predictions.

In principle, it is possible to calculate what has currently been 
measured, i.e.,
all charged hadrons in the trigger and other particles in the jet. 
In practice,
the task would be dauntingly complicated and involve many more terms than
what we have already considered, including thermal $s$ partons whose
parametrizations have been less reliably determined.  Besides, the current
data are in a passing phase; better particle identification is forth-coming.
What we have calculated are for clean triggers and associated particles, and
can be effectively compared with future data.

It is evident from the results of our study that the dihadron correlation is
dominated by the components that involve the highest number of thermal
partons.  The recombination mechanism boosts the yield when high-density
thermal partons are included, but also boosts the $p_T$ of the product when
the semi-hard shower partons are involved.  The effect has already been shown
to operate in the single-particle distributions, but now exhibits itself more
conspicuously in dihadron correlations in jets, since at least four partons are
involved, two of which can be thermal.  The difference between jets produced
in heavy-ion collisions compared to those produced in $pp$ collisions is huge,
as evidenced by the large ratio shown by the solid line in Fig.\ 3 
and by the ratios of the moments in Table II -- about 5 between
the first and third rows of values.

It is hard to see how the properties of the dihadron correlations in 
jets that we
have found can be reproduced in any fragmentation model even with medium
modification of the fragmentation function, whose focus has been on 
the effect of energy loss  \cite{whs}.  There are terms in our
recombination formulas, when the shower parton momenta are symmetrized,
that cannot be written in factorizable forms involving products of two FF's.
Leaving aside such technical details, let us accept the conceivable possibility
that perturbative branching of a hard parton can generate hard shower
partons, and that they can further fragment by means of suitable modified
FF's that mimic ${\cal TS}$ recombination.  If it is successful in the end, it
seems that the scheme would have lost the original advantage of the
fragmentation approach that relies on the universality of the FF's.  If the
modification of a FF depends sensitively on the detailed properties of the
partonic medium, then, by comparison, ${\cal TS}$ recombination would seem
to be a more direct and physically cogent approach to hadronization. 
Besides, as pointed out earlier, energy loss is not the issue in 
dihadron correlation.

There are limitations to the formalism that we have used for our 
calculations.
We have not considered the $Q^2$ dependence of dihadron 
correlation, since the SPD's used are for $Q$ fixed at 10 GeV/c 
\cite{hy5}. That limitation is not a matter of principle, but of 
practice. To account for the $Q^2$ evolution of the SPD's is a 
worthwhile problem in its own right, inasmuch as the same problem for 
the FF's has been pursued for decades \cite{bkk,sk}. To apply that 
evolutionary property to the dihadron correlation would be 
prohibitively complicated. Attempts have been initiated to 
investigate the evolutionary aspect of dihadron distribution of the 
fragmentation process in the operator formalism \cite{mw}. An input 
on the initial distribution for such an evolution would still have to 
involve the type of consideration presented here. Furthermore, how 
the thermal partons are to be incorporated in that approach is not 
clear.

 Another limitation is rooted in our formalism.
Since we have relied entirely on our 1D formulation of recombination, it is
not possible in the same formalism to address the question of angular
correlation in a jet.  Since jets are 3D objects, it is obvious that
longitudinal-transverse correlation within a jet can contain information about
the properties of the recombining partons that we cannot probe in the
simplified 1D formalism.  Clearly, despite the substantial progress that has
been made in our line of investigation, there remains much ground to 
improve and generalize in the study of the physics of hadronization 
in heavy-ion collisions.

\section*{Acknowledgment}
We are grateful to  X.\ N.\ Wang, A.\ Majumder, F.\ Wang, A.\ Sickles 
and M. Tannenbaum for helpful discussions and communication.
This work was supported, in part,  by the U.\ S.\ Department of Energy under
Grant No. DE-FG02-96ER40972  and by the Ministry of Education of China
under Grant No. 03113.

\newpage
\centerline{\large{\bf Figure Captions}}
\vskip0.5cm
\begin{description}

\item Fig.\ 1. Transverse momentum distribution of $\pi^+$ associated 
with a $\pi^+$ trigger. The contribution from terms of the form $\cal 
(TS)(TS)$ is shown in dashed line, while the contribution from terms 
of the form $\cal (TS)(SS)$ is shown in dashed-dot line. The solid 
line is the sum of all components. 

   \item Fig.\ 2. Associated particle ($\pi^+$) distribution with 
$\pi^+$ trigger for central Au+Au (solid), central d+Au (dashed) and 
peripheral d+Au collisions (dash-dot line).

\item Fig.\ 3. The ratio of central Au+Au to peripheral d+Au 
collisions (solid line) and that of central d+Au to peripheral d+Au 
collisions (dashed line).

\item Fig.\ 4. Transverse momentum distributions  of $\pi^+$, 
$\pi^-$ and $p$, associated with a $\pi^+$ trigger. The data are from 
STAR \cite{fq} for all charged hadrons in the trigger and associated 
particles.

\item Fig.\ 5. Transverse momentum distributions  of 
$\pi^+$ and $\pi^-$ (lowered by a factor of 2) associated with proton 
trigger.
\end{description}
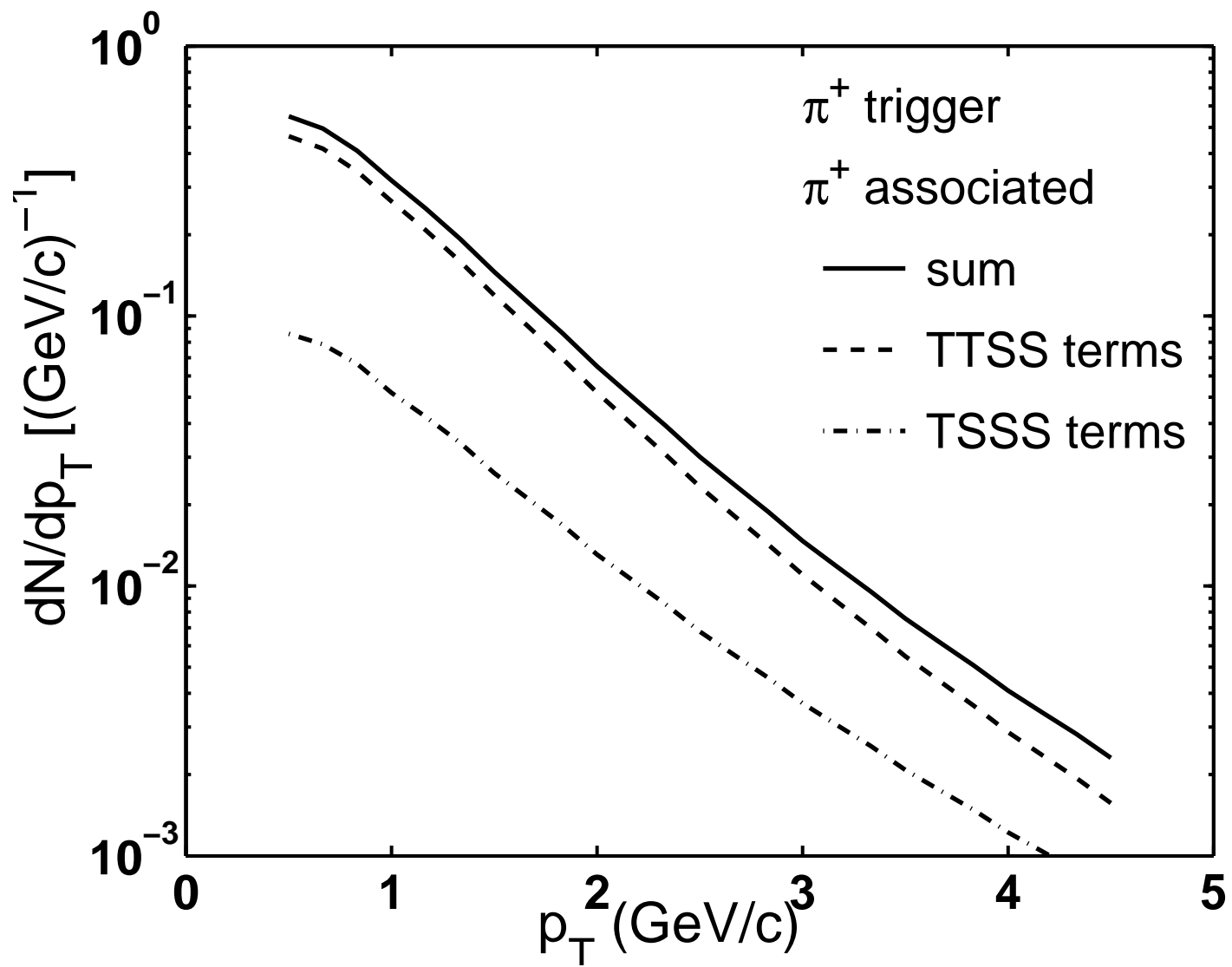
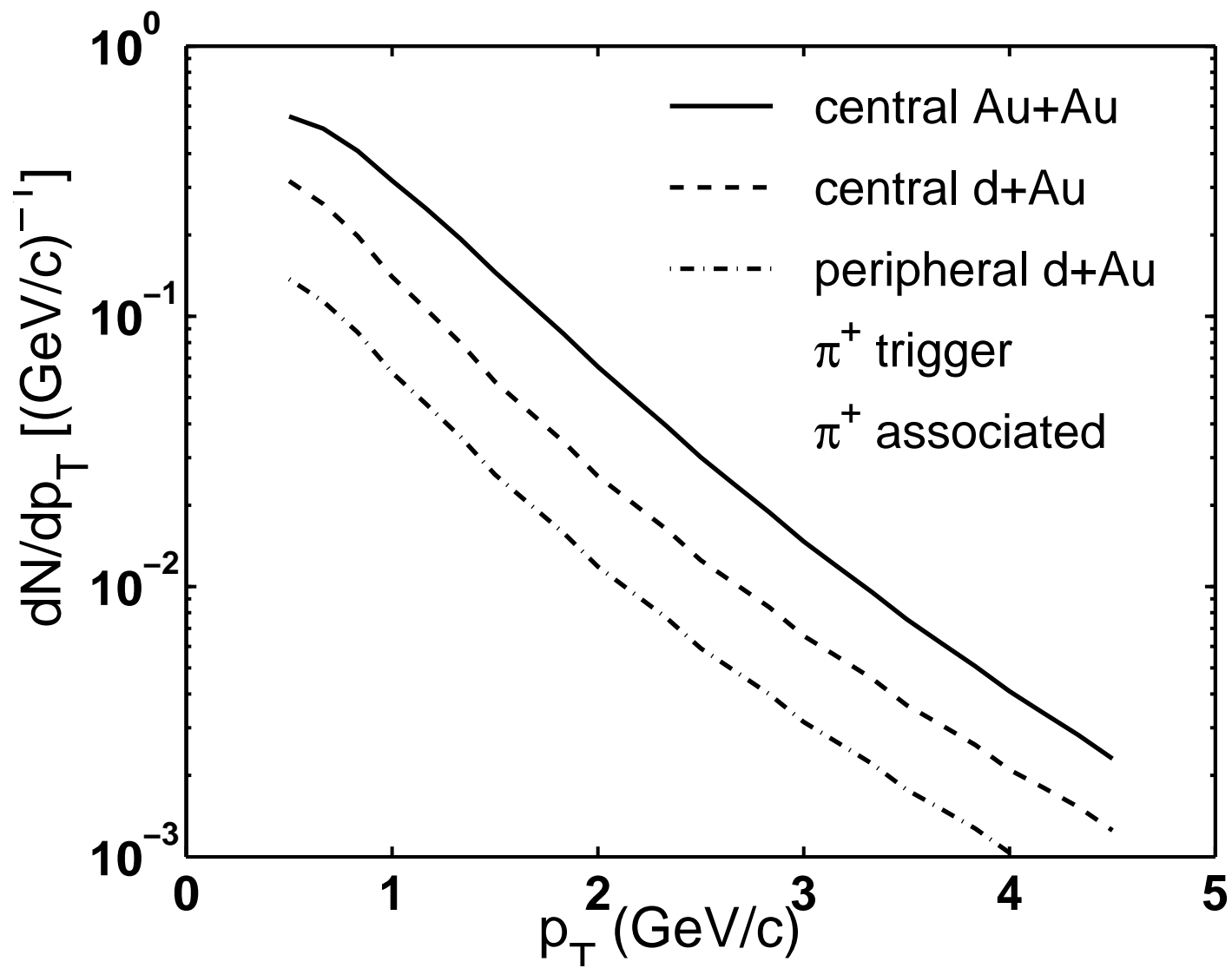
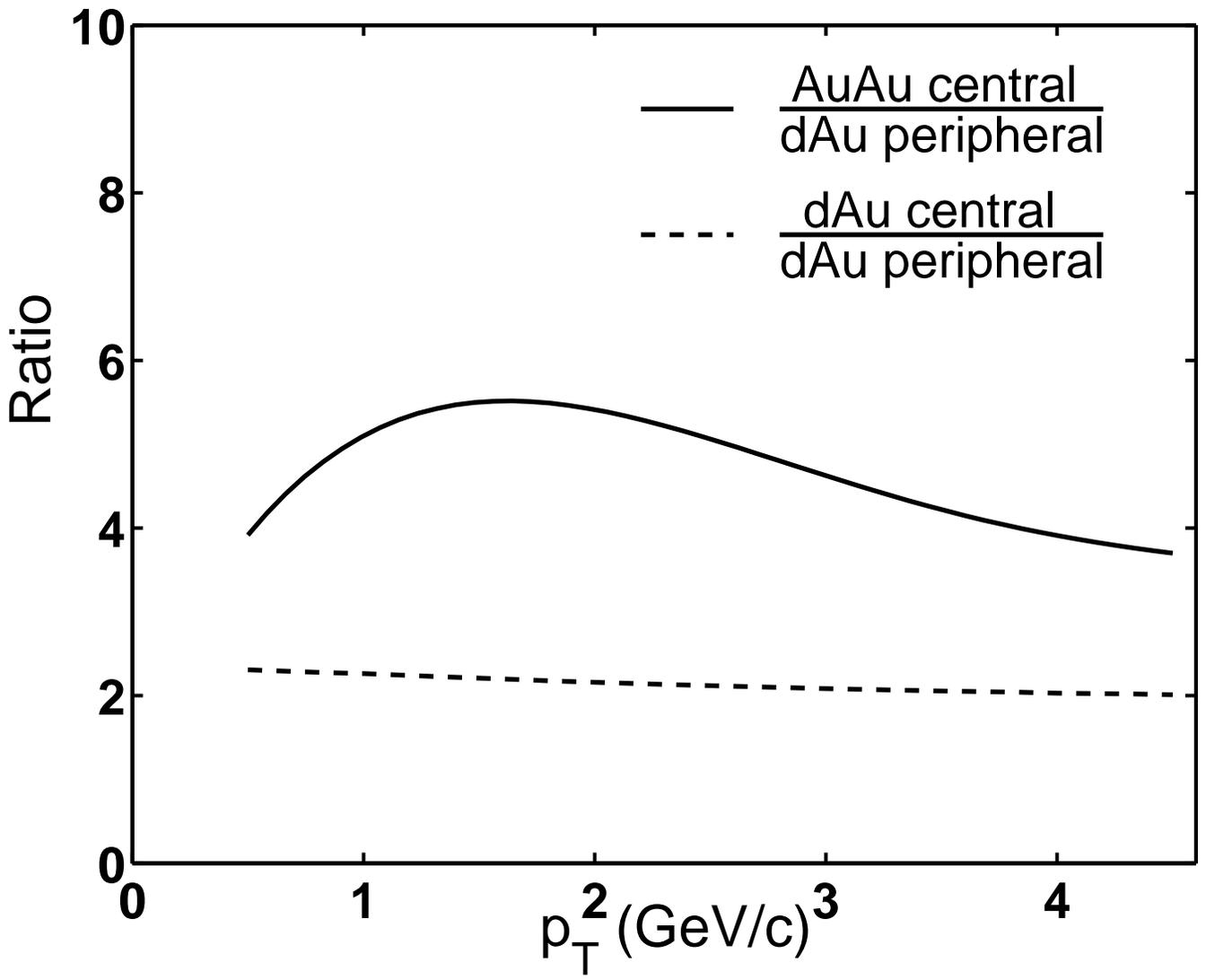
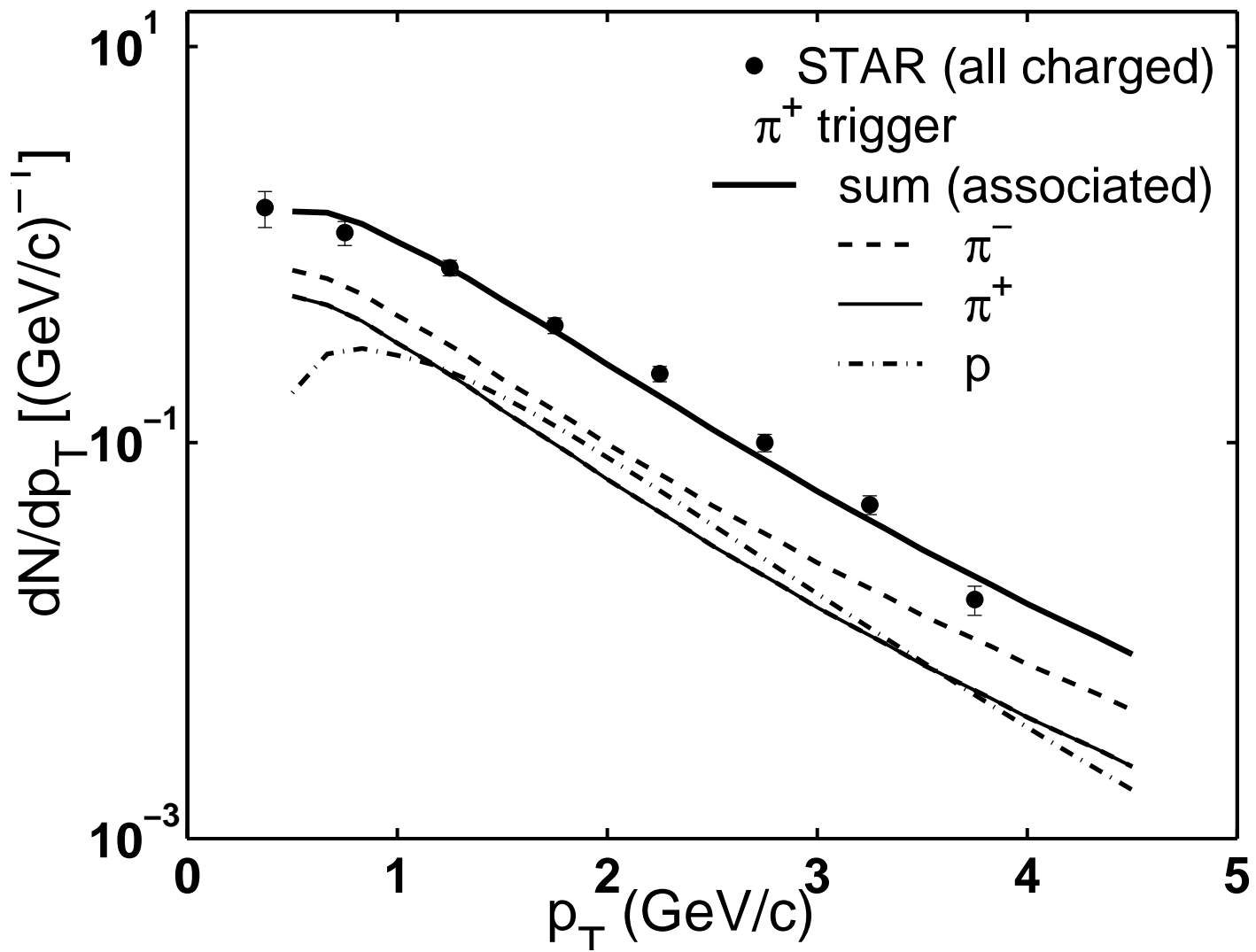
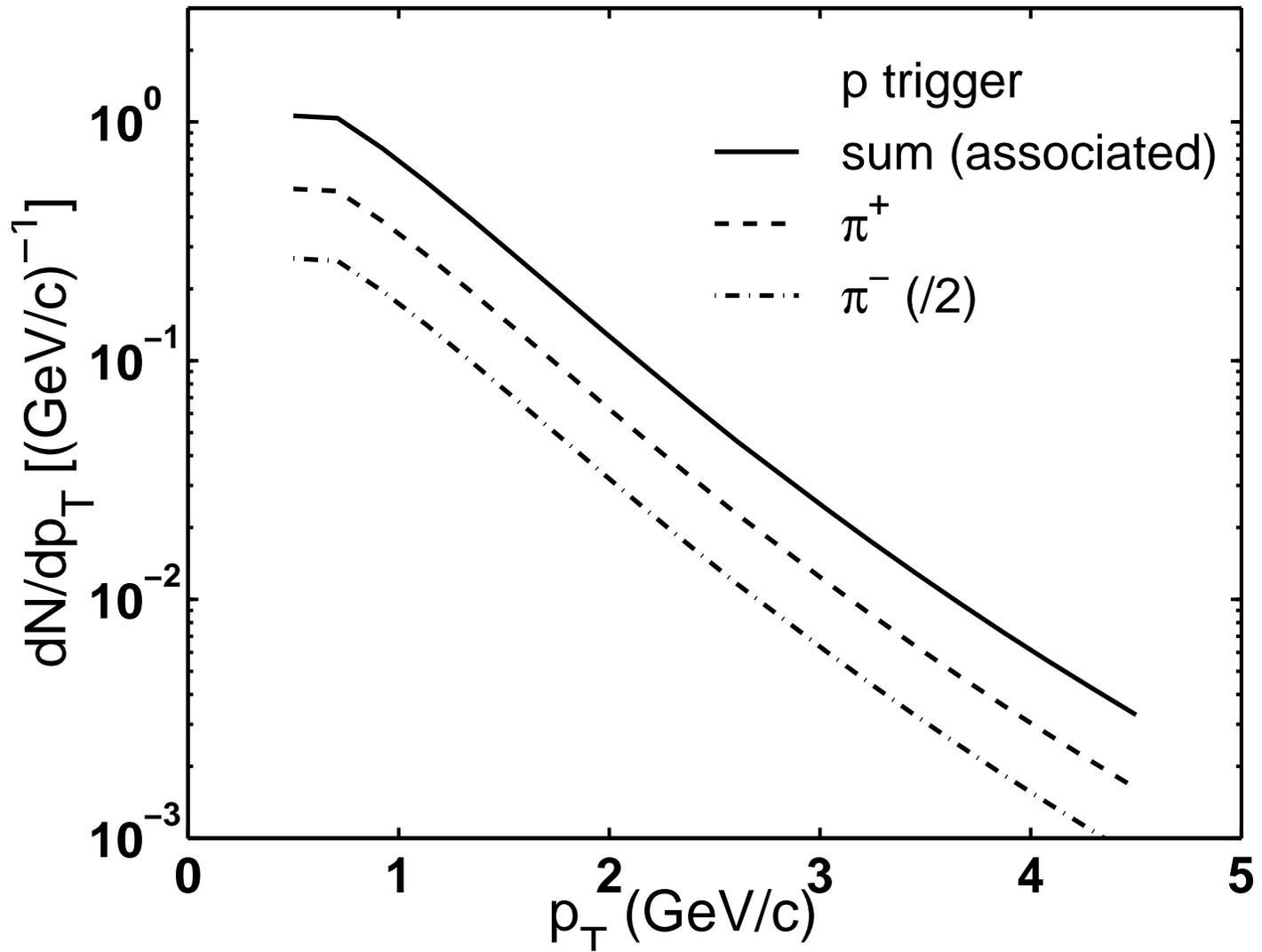
\end{document}